\documentstyle [12pt]{article}
\let\tilde=\widetilde

\def\tr{{\rm tr}}
\def\qdet{\hbox{q-det}}

\def\one#1{#1^{\raise5pt\hbox{$\scriptstyle\!\!\!\!1$}}\,{}}
\def\two#1{#1^{\raise5pt\hbox{$\scriptstyle\!\!\!\!2$}}\,{}}
\def\three#1{#1^{\raise5pt\hbox{$\scriptstyle\!\!\!\!3$}}\,{}}

\def\phi{\varphi}

\def\a{\alpha}
\def\b{\beta}
\def\d{\delta}
\def\D{\Delta}

\def\L{\Lambda}

\def\beq{\begin{equation}}
\def\eeq{\end{equation}}
\def\be{\begin{displaymath}}
\def\ee{\end{displaymath}}
\def\bm{\left(\begin{array}}
\def\em{\end{array}\right)}
\def\bds{\begin{description}}
\def\eds{\end{description}}
\def\?{(?)\marginpar{|?}}

\def\half{\frac{1}{2}}

\newtheorem{pr}{Proposition}

\newtheorem{lemma}{Lemma}
\newfont{\bbd}{msbm10 scaled\magstep1} %% if mbsm font not available use
\def\C{\hbox{\bbd C}}                  %% \def\C{{\bf C}}
\def\P{\hbox{\bbd P}}                  %% \def\P{{\cal P}}
\def\T{${\cal T}$}
\def\minor#1#2#3#4{\bigl[\vphantom{T}^{#1#2}_{#3#4}\bigr]}
\def\g{\gamma}
\def\ttb{\hat B}
\def\id{\hbox{{1}\kern-.25em\hbox{l}}}
\begin{document}
\title{Separation of Variables in the Quantum\\
       Integrable Models Related  \\
to the Yangian ${\cal Y}[sl(3)]$}
\author{E. K. Sklyanin \thanks{On leave from Steklov Mathematical Institute,
 Fontanka 27, St.Petersburg 191011, Russia.}\\
Isaac Newton Institute for Mathematical Sciences,\\
Cambridge University, Cambridge, CB3 0EH, U.K.
        }
\date{December 8, 1992}
\maketitle
\vskip-9.5cm
\hskip13.0cm
\sf NI-92013 \rm
\vskip9.5cm

{\bf Abstract.} There being no precise definition of the quantum
integrability, the separability of variables can serve as its
practical substitute.  For any quantum integrable model
generated by the Yangian ${\cal Y}[sl(3)]$ the canonical coordinates
and the conjugated operators are constructed which satisfy the
``quantum characteristic
equation'' (quantum counterpart of the spectral algebraic curve for
the L operator). The coordinates constructed provide a local separation
of variables. The conditions are enlisted which are necessary for
the global separation of variables to take place.

\section{Introduction}

Despite the huge body  of papers on ``quantum integrable
models'' written in last decades, ironically enough, there seems still
to be no satisfactory definition of quantum integrability \cite{Weigert}.
In the
classical case Liouville's definition of complete integrability
allows immediately to integrate the equations of motion in quadratures.
In the quantum case, on the contrary,
the mere  existence of commuting Hamiltonians provides no help in finding
their common spectrum and eigenfunctions which is a natural analog of
integrating equations of motion in the classical mechanics. The discrepancy
seems to be due to the difficulty with the concept of independent integrals
of motion in the quantum case \cite{Weigert}.

There exists, however, another concept in the classical mechanics which is
more or less equivalent to the integrability, namely, the separation of
variables in the Hamilton-Jacobi equation. Its advantage is that it has
a direct quantum counterpart.

Suppose that a quantum mechanical system
possesses a finite number
$D$ of commuting Hamiltonians $H_j$ ($j=1,\ldots,D$).
Suppose also that one can introduce $D$ pairs of canonically commuting
operators $x_j$, $p_j$
$$ [x_j,x_k]=[p_j,p_k]=0 \qquad [p_j,x_k]=-i\d_{jk} $$
such that:
\begin{enumerate}
\item The common spectrum of $\{x_j\}_{j=1}^D$ is simple, that is the whole
Hilbert space of quantum-mechanical states is isomorphic to a space of
functions on ${\rm spec}\{x_j\}_{j=1}^D$. The momenta $p_j$ are then realized
as the differentiations $p_j=-i\partial/\partial x_j$.
The simplicity of ${\rm spec}\{x_j\}_{j=1}^D$ replaces here the classical
concept of a Hamiltonian system having $D$ degrees of freedom.
\item There exist such polynomials $\Phi_j$ of $D+2$ variables that
\beq
\Phi_j(p_j,x_j,H_1,H_2,\ldots,H_D)=0 \qquad j=1,\ldots,D
\label{eq:sepvar}
\eeq
\end{enumerate}

The noncommuting operators in (\ref{eq:sepvar}) are assumed to be
ordered precisely in the same way they are enlisted, that is
$p_jx_jH_1H_2\ldots H_D$.

Now, let $\Psi(x_1,...,x_D)$ be a common eigenfunction of the Hamiltonians
$H_j$
\beq
H_j\Psi=h_j\Psi
\label{eq:Hpsi}
\eeq

Applying  the operator identities (\ref{eq:sepvar}) to
$\Psi$ and using (\ref{eq:Hpsi}) and the ordering of $H$'s in
(\ref{eq:sepvar}) one obtains for $\Psi$ the set of differential
equations
\beq
 \Phi_j(-i\frac{\partial}{\partial x_j},x_j,h_1,h_2,\ldots,h_D)
      \Psi(x_1,\ldots,x_D)=0 \qquad j=1,\ldots,D
\label{eq:PhiPsi}
\eeq
which obviously allows the separation of variables.
\beq
 \Psi(x_1,\ldots,x_D)=\prod_{j=1}^D \psi_j(x_j)
\label{eq:Psi}
\eeq

The original multidimensional spectral problem is thus reduced to the set
of one-dimensional multiparameter spectral problems
\beq
 \Phi_j(-i\frac{\partial}{\partial x_j},x_j,h_1,h_2,\ldots,h_D)
      \psi_j(x_j)=0 \qquad j=1,\ldots,D
\label{eq:Phipsi}
\eeq

The functions $\Phi_j$ being polynomials, the separated equations
(\ref{eq:Phipsi}) are ordinary differential equations.
More generally, one can consider $\Phi_j$
as symbols of pseudo\-differential operators. If one
allows, for instance,  $\Phi_j$ to depend on $p_j$ exponentially,
then (\ref{eq:Phipsi}) become finite-difference equations.

The above argument has, however, only a heuristic value since
it establishes only a local separation of variables and for the actual
(global) s.\ o.\ v.\  some more conditions are to be satisfied
(see Section 4).

It is tempting to adopt the separability of variables as a practical
definition of the quantum integrability. To this end, it is necessary to show
at least that the models generally referred to as ``quantum integrable'' ones
do allow the separation of variables in the above sense. First of all, it
concerns the spin chains  soluble via all variants of Bethe ansatz technique
which have no apparent resemblance to the separation of variables.
It is not until recently that an s.~o.~v.\  has been
constructed for the models generated by the Yangian ${\cal Y}[sl(2)]$, see
\cite{Skl:30,Skl:32} and the references therein.

The next natural step is to study the ${\cal Y}[sl(3)]$ case.
The relevant  quantum integrable models include $SU(3)$-invariant
spin chains \cite{Suth:PRB,KR:LOMI,KR:JPA} together with their
degenerated case (Gaudin model) \cite{Gaudin:book,Jurco:1,Jurco:2},
three-wave system (Lee model) \cite{KR:JPA},
$SU(2)$ vector Nonlinear Schr\"odinger equation \cite{Yang,Kulish:DAN},
all of them  well studied
via coordinate Bethe ansatz \cite{Suth:PRB,Gaudin:book,Yang},
algebraic BA \cite{KR:JPA,Jurco:1,Jurco:2,Kulish:DAN}
and analytic BA \cite{KR:LOMI} techniques.

The separation
of variables for the classical integrable $SL(3)$ magnetic chain was
constructed in \cite{Skl:34}, see also \cite{Adams} for the classical
$SL(N)$ case.
The present paper is the first in the series
devoted to the quantum $SL(3)$ case. It outlines a general scheme
allowing to construct a separation of variables for the models in question.
Intending to make the argument as general as possible we collect here
only the results which are common for various representations of
${\cal Y}[sl(2)]$. An implementation and adjustment  of the
presented scheme for the diverse particular models is left for
subsequent publications.

Using the experience acquired during the study of the quantum $SL(2)$ case
and the classical $SL(3)$ case we construct the variables
$x_j$ as the operator zeroes of certain operator polynomial $B(u)$ having
commuting coefficients.
In the classical case the exponents of the conjugated momenta $X_j=e^{p_j}$
are known to be the eigenvalues of the L operator $T(u)$ taken at $u=x_j$
and, as such, satisfy the corresponding characteristic   equation. We show that
the corresponding quantum variables $X_j$ also satisfy
a sort of ``quantum characteristic equation''
which fits the form (\ref{eq:sepvar}) and thus provides,
in principle, a separation of variables. The actual separation of
variables can be established, however, only if the
representation of the algebra formed by $X_jx_j$ satisfies several
additional conditions which we conjecture to be satisfied for
any representation of ${\cal Y}[sl(3)]$.

The conjectured separated equations are 3rd order finite-difference
equations. We show that our results agree with those obtained by means
of the algebraic Bethe ansatz.

\section{Description of the algebra}

Consider the associative algebra \T\ generated by $9(M+1)$ generators
$T^\a_{\b,m}$ ($\a,\b\in\{1,2,3\}$, $m\in\{0,1,\ldots,M\}$) and the quadratic
relations which are most conveniently described in terms of the
polynomials $T^\a_\b(u)=\sum_{m=0}^M u^m T^\a_{\b,m}$:
\begin{eqnarray}
 \lefteqn{
 (u-v)T^{\a_1}_{\b_1}(u)T^{\a_2}_{\b_2}(v)
+\eta T^{\a_2}_{\b_1}(u)T^{\a_1}_{\b_2}(v)} \nonumber \\
&& =(u-v)T^{\a_2}_{\b_2}(v)T^{\a_1}_{\b_1}(u)
+\eta T^{\a_2}_{\b_1}(v)T^{\a_1}_{\b_2}(u) \qquad \forall u,v
\label{eq:TT}
\end{eqnarray}
or, equivalently,
$$\sum_{\beta_1,\beta_2=1}^3
   R^{\alpha_1\alpha_2}_{\beta_1\beta_2}(u-v)
   T^{\beta_1}_{\gamma_1}(u)T^{\beta_2}_{\gamma_2}(v)
=\sum_{\beta_1,\beta_2=1}^3
 T^{\alpha_2}_{\beta_2}(v)T^{\alpha_1}_{\beta_1}(u)
 R^{\beta_1\beta_2}_{\gamma_1\gamma_2}(u-v)$$
where
\be
 R^{\a_1\a_2}_{\b_1\b_2}(u)=u\d^{\a_1}_{\b_1}\d^{\a_2}_{\b_2}
                       +\eta\d^{\a_1}_{\b_2}\d^{\a_2}_{\b_1}
\ee

Introducing the $3\times3$ matrix
$$ T(u)=\left(\begin{array}{ccc}
         T^{1}_{1}(u)&T^{1}_{2}(u)&T^{1}_{3}(u)\\
         T^{2}_{1}(u)&T^{2}_{2}(u)&T^{2}_{3}(u)\\
         T^{3}_{1}(u)&T^{3}_{2}(u)&T^{3}_{3}(u)
        \end{array}\right),                           $$
the unit operator $\id$ in $\C^3$ and
the permutation operator
$\P_{12}$ in the tensor product $\C^3\otimes\C^3$
$$ \id x=x \qquad \P_{12} x\otimes y =y\otimes x\qquad \forall x,y\in\C^3,$$
and the notation
$$ \one T\equiv T\otimes\id\qquad
   \two T\equiv\id\otimes T$$
one can rewrite (\ref{eq:TT}) in a compact form:
\beq
R(u-v)\one T(u)\two T(v)=\two T(v)\one T(u)R(u-v)
\label{eq:RTT}
\eeq
where
\beq
  R(u)=u\id\otimes\id+\eta\P_{12}
\label{eq:RXXX}
\eeq
is the well-known $SL(3)$-invariant solution to the Yang-Baxter
equation
\beq
  R_{12}(u)R_{13}(u+v)R_{23}(v)=
  R_{23}(v)R_{13}(u+v)R_{12}(u)
\label{eq:YBE}
\eeq
or, at length,
$$\sum_{\beta_1,\beta_2,\beta_3=1}^3
  R^{\alpha_1\alpha_2}_{\beta_1\beta_2}(u)
  R^{\beta_1\alpha_3}_{\gamma_1\beta_3}(u+v)
  R^{\beta_2\beta_3}_{\gamma_2\gamma_3}(v)
=\sum_{\beta_1,\beta_2,\beta_3=1}^3
  R^{\alpha_2\alpha_3}_{\beta_2\beta_3}(v)
  R^{\alpha_1\beta_3}_{\beta_1\gamma_3}(u+v)
  R^{\beta_1\beta_2}_{\gamma_1\gamma_2}(u)$$

It follows from the relations (\ref{eq:TT}) that the leading coefficients
$T^\a_{\b,M}$ of the polynomials $T^\a_\b(u)$ belong to the center of
the algebra \T. We shall suppose that $Z\equiv T_M$ is a  number  matrix
having distinct nonzero eigenvalues. Subsequently, some additional
nondegeneracy conditions will be imposed on $Z$.

In what follows we shall consider $T^\a_{\b,m}$ as linear operators
belonging to a representation of \T in a linear space $W$.
The representation theory for the algebra \T in case $Z^a_\b=\d^a_b$
is essentially equivalent  to that of the Yangian ${\cal Y}[sl(3)]$,
cf.\ \cite{KirR:LMP,Ch:,CP}.
For our purposes it is more convenient to take
a generic matrix $Z$ rather then unit one, which is equivalent  to
taking $GL(3)\otimes{\cal Y}[sl(3)]$ instead of ${\cal Y}[sl(3)]$.

We conclude this section with a synopsis of properties of quantum
minors and determinants \cite{KR:LOMI,KirR:LMP,Skl:11}.

Let $P^-_{12}$ and $P^-_{123}$ be the antisymmetrizers in
$\C^3\otimes\C^3$ and $\C^3\otimes\C^3\otimes\C^3$, respectively,
\begin{eqnarray}
 P^-_{12}&=&\half(\id\otimes\id-\P_{12}) \nonumber \\
 P^-_{123}&=&\frac{1}{6}(\id\otimes\id\otimes\id+\P_{12}\P_{23}+\P_{23}\P_{12}
          -\P_{12}-\P_{13}-\P_{23})  \nonumber
\end{eqnarray}
and $\tr_j$ stand for the trace over the $j$-th copy of $\C^3$
in $\C^3\otimes\C^3\otimes\C^3$.

The quantum determinant $d(u)$ of $T(u)$ is then defined as
\begin{eqnarray}
   d(u)=\qdet T(u)&=&
  \tr_{123}P_{123}^-\one T(u)\two T(u+\eta)\three T(u+2\eta) \nonumber
\end{eqnarray}
and generates, together with $T_M$, the center of the algebra \T.
We shall assume that the coefficients of $d(u)$, like $T^\a_{\b,M}$
are numbers.

The quantum minor $T^{\a_1\a_2}_{\b_1\b_2}(u)$ is defined as the
quantum determinant of the $2\times2$ submatrix formed by the
rows $\a_1<\a_2$ and the columns $\b_1<\b_2$ of the matrix $T(u)$
\begin{eqnarray}
 T\minor{\a_1}{\a_2}{\b_1}{\b_2}(u)&=&
     \qdet\bm{cc}
      T^{\a_1}_{\b_1}(u) & T^{\a_1}_{\b_2}(u) \nonumber \\
      T^{\a_2}_{\b_1}(u) & T^{\a_2}_{\b_2}(u)
     \em \\
&=&T^{\a_2}_{\b_2}(u)T^{\a_1}_{\b_1}(u+\eta)-T^{\a_1}_{\b_2}(u)T^{\a_2}_{\b_1}
(u+\eta) \nonumber \\
&=&T^{\a_1}_{\b_1}(u)T^{\a_2}_{\b_2}(u+\eta)-T^{\a_2}_{\b_1}(u)T^{\a_1}_{\b_2}
(u+\eta) \nonumber \\
&=&T^{\a_1}_{\b_1}(u+\eta)T^{\a_2}_{\b_2}(u)-T^{\a_1}_{\b_2}(u+\eta)
T^{\a_2}_{\b_1}(u) \nonumber \\
&=& T^{\a_2}_{\b_2}(u+\eta)T^{\a_1}_{\b_1}(u)-T^{\a_2}_{\b_1}(u+\eta)
T^{\a_1}_{\b_2}(u)
\label{eq:minors}
\end{eqnarray}

Consider the matrix $U(u)$ formed by the quantum minors
\beq
 U(u)=\bm{ccc}
 \phantom{-}T\minor 2323(u) & -T\minor 2313(u) & \phantom{-}T\minor 2312(u) \\
       -T\minor 1323(u) & \phantom{-}T\minor 1313(u) & -T\minor 1312(u) \\
 \phantom{-}T\minor 1223(u) & -T\minor 1213(u) & \phantom{-}T\minor 1212(u)
      \em
\label{eq:defU}
\eeq

An equivalent expression for $U(u)$ is given by
$$ \three U(u)^t=3 \tr_{23}P^-_{123}\one T(u)\two T(u+\eta)
$$
where $t$ stands for the matrix transposition.

The matrix $U(u)$ allows to invert $T(u)$:
\begin{eqnarray}
d(u)&=& T(u)^tU(u+\eta)=U(u+\eta)T(u)^t \nonumber \\
    &=& U(u)^tT(u+2\eta)=T(u+2\eta)U(u)^t
\end{eqnarray}
or, in expanded form,
\begin{eqnarray}
d(u)\d_{\a\g}&=& \sum_\b T^{\b}_{\a}(u)U^{\b}_{\g}(u+\eta)=
 \sum_\b U^{\a}_{\b}(u+\eta)T^{\g}_{\b}(u) \nonumber \\
&=& \sum_\b U^{\b}_{\a}(u)T^{\b}_{\g}(u+2\eta)=
 \sum_\b T^{\a}_{\b}(u+2\eta)U^{\g}_{\b}(u)
\label{eq:TU}
\end{eqnarray}

The commutation relations for $T(u)$ and $U(u)$,
like (\ref{eq:RTT}), can also be written in the R matrix form
($\hat R(u)=[R(u+\eta)^{-1}]^{t_2}$)
\beq
 \hat R(u-v)\one T(u)\two U(v)=\two U(v)\one T(u)\hat R(u-v)
\eeq
\beq
 R(u-v)\one U(u)\two U(v)=\two U(v)\one U(u)R(u-v)
\eeq
or, at length,
\begin{eqnarray}
 \lefteqn{
 (u-v+\eta)T^{\a_1}_{\b_1}(u)U^{\a_2}_{\b_2}(v)
-\eta\d_{\a_1\a_2}\sum_\g T^{\g}_{\b_1}(u)U^{\g}_{\b_2}(v)} \nonumber \\
&& =(u-v+\eta)U^{\a_2}_{\b_2}(v)T^{\a_1}_{\b_1}(u)
-\eta\sum_\g U^{\a_2}_{\g}(v)T^{\a_1}_{\g}(u)\d_{\b_1\b_2}
\label{eq:ortTU}
\end{eqnarray}

\begin{eqnarray}
 \lefteqn{
 (u-v)U^{\a_1}_{\b_1}(u)U^{\a_2}_{\b_2}(v)
+\eta U^{\a_2}_{\b_1}(u)U^{\a_2}_{\b_1}(v)} \nonumber \\
&& =(u-v)U^{\a_2}_{\b_2}(v)U^{\a_1}_{\b_1}(u)
+\eta U^{\a_2}_{\b_1}(v)U^{\a_1}_{\b_2}(u)
\label{eq:UU}
\end{eqnarray}

The quantum minors of $U(u)$ are, in turn,  expressed in terms of $T(u)$
\beq
 U\minor{\a_1}{\a_2}{\b_1}{\b_2}(u)=s_\a s_\b T^{\a_3}_{\b_3}(u+\eta)d(u)
\label{eq:minU}
\eeq
where the triplets $(\a_1\a_2\a_3)$ and $(\b_1\b_2\b_3)$ are permutations of
$(123)$
and $s_\a$ and $s_\b$ are the corresponding signatures.

The commuting Hamiltonians are given by the matrix traces of $T(u)$
and $U(u)$
\beq
\begin{array}{rcl}
 t_1(u)&=&\tr T(u)\equiv\sum_\a T^{\a}_{\a}(u) \\
 t_2(u)&=&\tr U(u)=\tr_{12}P^-_{12}\one T(u) \two T(u+\eta)
\end{array}
\eeq
which are operator-valued polynomials of degree $M$ and $2M$,
respectively.

$$ [t_1(u),t_1(v)]=[t_1(u),t_2(v)]=[t_2(u),t_2(v)]=0 $$

Due to the assumption made that the leading coefficient of the
polynomial $T(u)$ is a nondegenerate number matrix the total
number of commuting Hamiltonians is $M+2M=3M$, cf. \cite{Skl:34}.

\section{Construction of canonical variables}

The commuting Hamiltonians being described, the next step is to
construct the separated variables $x_n$. In the classical case \cite{Skl:34}
they are constructed as the zeroes of the polynomial
$B(u)=T^2_3(u)U^3_1(u)-T^1_3(u)U^3_2(u)$ of degree $3M$.
In the quantum case, let us define the quantum operator-valued polynomial
$B_c(u)$ as
\beq
 B_c(u)=T^{2}_{3}(u)U^{3}_{1}(u-c)-T^{1}_{3}(u)U^{3}_{2}(u-c)
\label{eq:defBc}
\eeq

The parameter $c$ is the anticipated quantum correction which will be
fixed in the next section. The results of the present section are valid
for any value of $c$.

\begin{pr}
 $B_c(u)$ is a commutative family of operators
\beq
  [B_c(u),B_c(v)]=0
\label{eq:commB}
\eeq
\label{commB}
\end{pr}
The proof is given by a direct calculation. One takes the product
$B_c(u)B_c(v)$ and, after substituting the expressions (\ref{eq:defBc}) for
$B_c$ and expanding the brackets, tries to bring the $u$-dependent terms
through the $v$-dependent ones. To this end one notices that
the nontrivial commutations occur only inside the pairs
($T^{2}_{3}$, $T^{1}_{3}$) and ($U^{3}_{1}$, $U^{3}_{2}$)
and applies the relations
\beq
 T^{\a}_{3}(u)T^{\b}_{3}(v)=\frac{u-v}{u-v-\eta}T^{\b}_{3}(v)T^{\a}_{3}(u)
                    -\frac{\eta}{u-v-\eta}T^{\a}_{3}(v)T^{\b}_{3}(u)
\label{eq:TT3}
\eeq
\beq
 U^{3}_{\a}(u)U^{3}_{\b}(v)=\frac{u-v}{u-v+\eta}U^{3}_{\b}(v)U^{3}_{\a}(u)
                    +\frac{\eta}{u-v+\eta}U^{3}_{\a}(v)U^{3}_{\b}(u)
\label{eq:UU3}
\eeq
which follow from (\ref{eq:TT}) and (\ref{eq:UU}).
Finally, after multiple cancellations, one arrives at $B_c(v)B_c(u)$.

Let us assume now that the matrix $Z$ is such that the leading coefficient
${\cal B}$
of $B_c(u)$ is nonzero. It means that $Z$ should lie outside of certain
algebraic manifold of codimension 1 which must be true for a generic
matrix $Z$. We can define now the coordinate operators $x_n$
($n=1,\ldots,3M$) as
the zeroes of the polynomial $B_c(u)$ of degree $3M$.
\beq
 B(u)={\cal B}\prod_{n=1}^{3M}(u-x_n)\qquad
B(x_n)=0, \qquad n=1,\ldots,3M
\label{eq:defx}
\eeq

They commute
\beq
         [x_m,x_n]=0
\label{eq:xx}
\eeq
by virtue of (\ref{eq:commB}).
The detailed description of the
construction and the discussion of the mathematical subtleties of
handling the zeroes of operator-valued polynomials
are given in \cite{Skl:32} where the $SL(2)$ case is considered.
Since in the $SL(3)$ case the argument is the same, we do not reproduce
it here.

We shall assume that the common spectrum of $x_n$ is simple and
that the representation space $W$ is realized as some space of
functions on ${\rm spec}\{x_n\}_{n=1}^{3M}$. The precise description
of the functional space depends on the particular representation of
\T, see \cite{Skl:30} for discussion of possibilities in $SL(2)$ case.

The next step is the construction of the variables canonically
conjugated to $x_n$. Define the rational operator-valued function
$A_c(u)$ as
\beq
 A_c(u)=-T^2_3(u-\eta)^{-1}U^3_2(u-c)=-U^3_2(u-c)T^2_3(u-\eta)^{-1}
\label{eq:defAc}
\eeq

Note that the two expressions for $A_c(u)$ are equivalent since
$T^2_3(u)$ and $U^3_2(v)$ commute according to (\ref{eq:TU}).
Moreover, since $T^2_3(u)$ and $U^3_2(u)$ are themselves
commuting families, $A_c(u)$ is also a commuting family.
\beq
[A_c(u),A_c(v)]=0
\label{eq:commAA}
\eeq
\begin{pr}
 $A_c(u)$ and $B_c(v)$ satisfy the following commutation relation
\begin{eqnarray}
 \lefteqn{
 (u-v)A_c(u)B_c(v)-(u-v-\eta)B_c(v)A_c(u)} \nonumber \\
&&=\eta B_c(u)A_c(v)T^2_3(u-\eta)^{-1}T^2_3(u)^{-1}T^2_3(v-\eta)T^2_3(v)
\label{eq:commBA}
\end{eqnarray}
\label{commBA}
\end{pr}
{\bf Proof.} Notice that the following identity holds
\begin{eqnarray}
&& (u-v)U^3_2(u-c)B_c(v)T^2_3(u-\eta)-(u-v-\eta)T^2_3(u-\eta)B_c(v)U^3_2(u-c)
  \nonumber \\
&&\qquad=\eta \tilde B_c(u)T^2_3(v)U^3_2(v-c),
\label{eq:UBT}
\end{eqnarray}
where
\beq
 \tilde B_c(u)\equiv T^2_3(u-\eta)U^3_1(u-c)-T^1_3(u-\eta)U^3_2(u-c),
\eeq
which can be verified by a direct calculation in the same manner as in the
previous proof. Using the simple relation
\beq
 T^{\a}_{3}(u-\eta)T^{\b}_{3}(u)=T^{\b}_{3}(u-\eta)T^{\a}_{3}(u)
\label{eq:swapT}
\eeq
which follows from (\ref{eq:TT3}) one can verify also another identity
\beq
 \tilde B_c(u)T^2_3(u)=T^2_3(u-\eta)B_c(u)
\label{eq:swapB}
\eeq

Now, using (\ref{eq:swapB}) and the commutativity of $A_c(v)$ and $T^2_3(u)$,
we transform the right hand side of (\ref{eq:UBT}) into
\begin{eqnarray}
 \lefteqn{
 \eta T^2_3(u-\eta)B_c(u)T^2_3(u)^{-1}T^2_3(v)U^3_2(v-c) }  \nonumber\\
&&=-\eta T^2_3(u-\eta)B_c(u)A_c(v)T^2_3(u)^{-1}T^2_3(v)T^2_3(v-\eta)
\nonumber
\end{eqnarray}

Finally, multiplying the resulting identity by $T^2_3(u-\eta)^{-1}$ both from
the left and from the right, we arrive at (\ref{eq:commBA}).

In the classical case \cite{Skl:34} the exponents $X_n=e^{p_n}$
of the momenta
$p_n$ canonically conjugated to the coordinates $x_n$ are given
by the formula $X_n=A(x_n)$. In order to transfer the definition
to the quantum case it is necessary to describe what is understood
under the substitution of the operator values $u=x_n$ into $A_c(u)$.
Following \cite{Skl:32} we fix the operator ordering putting $x$'s
to the left. So, let for any rational operator-valued function
$F(u)$ the symbol $[F(u)]_u=x_n$ be defined as the Cauchy integral
\beq
 [F(u)]_{u=x_n}=\frac{1}{2\pi{\rm i}}
\int_\Gamma du (u-x_n)^{-1}F(u)
\label{eq:defsubx}
\eeq
taken over a closed  contour $\Gamma$ encircling counterclockwise
the spectrum of $x_n$ and
leaving outside  the poles of $F(u)$. Such a contour exists if
${\rm spec}x_n\cap{\rm poles}\{F(u)\}=\emptyset$. For the polynomials
$F(u)=\sum_pu^pF_p$ the definition $[F(u)]_u=x_n=\sum_px_n^pF_p$
given in \cite{Skl:32} is obviously recovered.

Define now $X_n$ as
\beq
 X_n=[A_c(u)]_{u=x_n}
\label{eq:defX}
\eeq

The condition
${\rm spec}\{x_n\}_{n=1}^{3M}\cap{\rm poles}\{A_c(u)\}=\emptyset$ should be
investigated for any concrete representation of the algebra \T,
and here we assume it to be satisfied.

The identities (\ref{eq:commAA}) and (\ref{eq:commBA}) proven above
entrain immediately the commutation relations
\beq
 [X_m,X_n]=0
\label{eq:XX}
\eeq
\beq
 X_mx_n=(x_n-\eta\d_{mn})X_m
\label{eq:Xx}
\eeq

The derivation of (\ref{eq:XX}) and (\ref{eq:Xx}) is the same as in
\cite{Skl:32}, so we do not reproduce it here.

\section{Quantum characteristic equation}

In the present paragraph we restrict the parameter $c$ to the value
$c=\eta$ so that now
\beq
 B(u)=T^2_3(u)U^3_1(u-\eta)-T^1_3(u)U^3_2(u-\eta)
\label{eq:defB}
\eeq

\beq
 A(u)=-T^2_3(u-\eta)^{-1}U^3_2(u-\eta)
     =-U^3_2(u-\eta)T^2_3(u-\eta)^{-1}
\label{eq:defA}
\eeq

\begin{pr}
 The following equality (``quantum characteristic equation'' for $X_n$)
holds
\beq
 X_n^3-X_n^2[t_1(u)]_{u=x_n}+X_n[t_2(u-\eta)]_{u=x_n}-d(x_n-2\eta)=0
\label{eq:characteristicX}
\eeq
\label{characteristicX}
\end{pr}

The identity (\ref{eq:characteristicX}) can be thought of as describing
the ``quantum algebraic spectral curve'' for the matrix $T(u)$.
It is an open question if the proposition \ref{characteristicX} can be
generalized for arbitrary value of $c$ and thus
if the variables $X_n$, $x_n$ provide a separation of coordinates
for some quantum integrable system.

The proof is based on the identity
\begin{eqnarray}
 \lefteqn{
 U^3_2(u-\eta)U^3_2(u-2\eta)U^3_2(u-3\eta) } \nonumber \\
&&+U^3_2(u-\eta)U^3_2(u-2\eta)T^2_3(u-3\eta)t_1(u-2\eta)  \nonumber \\
&&+U^3_2(u-\eta)T^2_3(u-2\eta)T^2_3(u-3\eta)t_2(u-2\eta) \nonumber \\
&&+T^2_3(u-\eta)T^2_3(u-2\eta)T^2_3(u-3\eta)d(u-2\eta) \nonumber \\
&&=\ttb(u)C(u)
\label{eq:UUU}
\end{eqnarray}
where
\beq
 \ttb(u)=T^2_3(u-3\eta)U^3_1(u-\eta)
-T^1_3(u-3\eta)U^3_2(u-\eta)
\label{eq:defttB}
\eeq
\beq
 C(u)=T^2_3(u-2\eta)U^{1}_{2}(u-2\eta)-T^{2}_{1}(u-2\eta)U^3_2(u-2\eta)
\label{eq:defC}
\eeq

Suppose for a moment that (\ref{eq:UUU}) is proven and
consider another identity
\beq
 \ttb(u)T^2_3(u-2\eta)T^2_3(u-\eta)T^2_3(u)=
T^2_3(u-\eta)T^2_3(u-2\eta)T^2_3(u-3\eta)B(u)
\label{eq:swaptB}
\eeq
which is derived in the same manner as (\ref{eq:swapB}).

Now let us multiply the equality (\ref{eq:UUU}) from the left by
${T^2_3(u-\eta)^{-1}}\*{T^2_3(u-2\eta)^{-1}}\*{T^2_3(u-3\eta)^{-1}}$, then use
the identity (\ref{eq:swaptB}) and the definition (\ref{eq:defA}) of $A(u)$.
The result is
\begin{eqnarray}
 \lefteqn{
 A(u)A(u-\eta)A(u-2\eta)-A(u)A(u-\eta)t_1(u-2\eta) } \nonumber \\
&&+A(u)t_2(u-2\eta)-d(u-2\eta) \nonumber \\
&&=-B(u)T^2_3(u-2\eta)^{-1}T^2_3(u-\eta)^{-1}T^2_3(u)^{-1}C(u)
\label{eq:AAA}
\end{eqnarray}

To obtain the desired characteristic equation (\ref{eq:characteristicX})
it suffices now to substitute $u=x_n$ into (\ref{eq:AAA}) from the left
and apply the following lemma.

\begin{lemma}
 For any operator-valued polynomial  $F(u)$
\beq
 [A(u)F(u-\eta)]_{u=x_n}=X_n[F(u)]_{u=x_n}
\label{eq:lemma}
\eeq
\label{lemma}
\end{lemma}

To conclude the proof it remains to prove the identity (\ref{eq:UUU})
and the lemma \ref{lemma}.

Expanding the right hand side of (\ref{eq:UUU}) and reordering
commuting factors we get four terms
\begin{eqnarray}
 \hat B(u)C(u)&=&\phantom{-}
U^3_2(u-\eta)U^3_2(u-\eta)T^1_3(u-3\eta)T^2_1(u-2\eta) \nonumber \\
&&-U^3_2(u-\eta)T^1_3(u-3\eta)T^2(u-2\eta)U^1_2(u-2\eta) \nonumber \\
&&-T^2_3(u-3\eta)U^3_1(u-\eta)U^3_2(u-2\eta)T^2_1(u-\eta) \nonumber \\
&&+T^2_3(u-3\eta)T^2_3(u-2\eta)U^3_1(u-\eta)U^1_2(u-2\eta)
\label{eq:BC}
\end{eqnarray}

Making in the first term substitution
$$T^1_3(u-3\eta)T^2_1(u-2\eta)=U^3_2(u-3\eta)
  +T^2_3(u-3\eta)T^1_1(u-2\eta), $$
see (\ref{eq:defU}) and (\ref{eq:minors}), and replacing $T^1_1$
with $t_1-T^2_2-T^3_3$ we obtain the expression
\begin{eqnarray}
&& U^3_2(u-\eta)U^3_2(u-2\eta)U^3_2(u-3\eta)
  +U^3_2(u-\eta)U^3_2(u-2\eta)T^2_3(u-3\eta)t_1(u-2\eta) \nonumber \\
&&\quad -U^3_2(u-\eta)U^3_2(u-2\eta)T^2_3(u-3\eta)
       [T^2_2(u-2\eta)+T^3_3(u-2\eta)] \nonumber
\end{eqnarray}

Analogously, in the fourth term of (\ref{eq:BC}) we make the substitution
$$ U^3_1(u-\eta)U^1_2(u-2\eta)=U^3_2(u-\eta)U^1_1(u-2\eta)
  +T^2_3(u-\eta)d(u-2\eta), $$
see (\ref{eq:minU}), and replace $U^1_1$ with $t_2-U^2_2-U^3_3$
obtaining
\begin{eqnarray}
&&T^2_3(u-\eta)T^2_3(u-2\eta)T^2_3(u-3\eta)d(u-2\eta)
 +U^3_2(u-\eta)T^2_3(u-2\eta)T^2_3(u-3\eta)t_2(u-2\eta) \nonumber \\
&&-U^3_2(u-\eta)T^2_3(u-2\eta)T^2_3(u-3\eta)
 [U^2_2(u-2\eta)+U^3_3(u-2\eta)] \nonumber
\end{eqnarray}

Notice that all the four terms of the left hand side of (\ref{eq:UUU})
are cancelled by the right hand side terms. Consider now the second term
in (\ref{eq:BC}) and apply the identity
$$ T^1_3(u-3\eta)T^2_3(u-2\eta)=T^2_3(u-3\eta)T^1_3(u-2\eta), $$
see (\ref{eq:swapT}), and then the identity
$$ T^1_3(u-2\eta)U^1_2(u-2\eta)=-T^2_3(u-2\eta)U^2_2(u-2\eta)
   -U^3_2(u-2\eta)T^3_3(u-2\eta), $$
see (\ref{eq:ortTU}). The result is
$$ U^3_2(u-\eta)T^2_3(u-3\eta)T^2_3(u-2\eta)U^2_2(u-2\eta)
  +U^3_2(u-\eta)U^3_2(u-2\eta)T^2_3(u-3\eta)T^3_3(u-2\eta) $$

Analogously, the third term of (\ref{eq:BC}) is transformed,
with the use of the identities
$$ U^3_1(u-\eta)U^3_2(u-2\eta)=U^3_2(u-\eta)U^3_1(u-2\eta) $$
and
$$ U^3_1(u-2\eta)T^2_1(u-2\eta)=-U^3_2(u-2\eta)T^2_2(u-2\eta)
  -T^2_3(u-2\eta)U^3_3(u-2\eta), $$
into
$$U^3_2(u-\eta)U^3_2(u-2\eta)T^2_3(u-3\eta)T^2_2(u-2\eta)
 +U^3_2(u-\eta)T^2_3(u-2\eta)T^2_3(u-3\eta)U^3_3(u-2\eta) $$

Collecting all the terms obtained we observe their total cancellation,
the identity (\ref{eq:UUU}) being thus proved.

To prove the lemma \ref{lemma} consider the left hand side of
(\ref{eq:lemma}) and apply the definition (\ref{eq:defsubx})
$$ [A(u)F(u-\eta)]_{u=x_n}=\frac{1}{2\pi{\rm i}}
\int_\Gamma du (u-x_n)^{-1}A(u)F(u-\eta)=\ldots
$$

Representing $F(u-\eta)$ as another Cauchy integral over the contour
$\Gamma^\prime$ encircling the point $u$ one rewrites
the previous expression as
$$ \ldots=\frac{1}{2\pi{\rm i}}\int_\Gamma du
          \frac{1}{2\pi{\rm i}}\int_{\Gamma^\prime} dv
(u-x_n)^{-1}(v-u)^{-1}A(u)F(v-\eta)=\ldots
$$

Changing the order of integrals and using the definition (\ref{eq:defX})
of $X_n$ and the commutation relation (\ref{eq:Xx})
one obtains finally the right hand side of (\ref{eq:lemma}).
\begin{eqnarray}
 \ldots&=&\frac{1}{2\pi{\rm i}}\int_{\Gamma^\prime} dv
(v-x_n)^{-1}X_nF(v-\eta)                \nonumber  \\
&=&\frac{1}{2\pi{\rm i}}\int_{\Gamma^\prime} dv
X_n(v-x_n-\eta)^{-1}F(v-\eta)                \nonumber  \\
 &=& X_n[F(v)]_{v=x_n} \nonumber
\end{eqnarray}

The proposition \ref{characteristicX} being thus proven, we can discuss
its corollaries.  The equation (\ref{eq:characteristicX}) obviously
fits the form (\ref{eq:sepvar}) since the operator ordering in
(\ref{eq:characteristicX}) is the same as postulated for (\ref{eq:sepvar}).
Following the heuristic argument given in the Introduction
one can expect that the quantum characteristic
equation (\ref{eq:characteristicX}) yields the separation of variables
for the $3M$ commuting Hamiltonians given by the
coefficients of the polynomials $t_1(u)$ and $t_2(u)$.
If $X_n$ are realized
as shift operators $X_n=\exp\{-\eta\partial/\partial x_n\}$
the resulting
separated equations (\ref{eq:Phipsi}) become the third order
finite difference equations
$$ \psi_n(x_n-3\eta)-\tau_1(x_n-2\eta)\psi_n(x_n-2\eta)
+\tau_2(x_n-2\eta)\psi_n(x_n-\eta)-d(x_n-2\eta)\psi(x_n)=0 $$
where $\tau_{1,2}(u)$ are the eigenvalues of the operators $t_{1,2}(u)$.

However, one cannot always take $X_n$
as pure shifts. Generally speaking, $X_n$ should contain
a cocycle factor $\D_n(x)$
$$ X_n\Psi(\ldots, x_n,\ldots)=
\D_n(x_1,\ldots,x_{3M})\Psi(\ldots,x_n-\eta,\ldots) $$

There is a liberty of
canonical transformations $\Psi(x)\rightarrow \rho(x)\Psi(x)$
and, respectively,
\beq
\D_n(\ldots,x_n,\ldots)\rightarrow
\frac{\rho(\ldots,x_n-\eta,\ldots)}{\rho(\ldots,x_n,\ldots)}
\D_n(\ldots,x_n,\ldots)
\label{eq:canon}
\eeq
where $\rho(x)$ is a nonzero function on ${\rm spec}\{x_n\}_{n=1}^{3M}$.

For the finite dimensional representations of \T\ the cocycles
$\D_n(x)$ certainly cannot be equivalent to the trivial ones $D_n(x)\equiv 1$
since they must have zeroes on the boundary of the finite set
 which cannot be removed by any
canonical transformation (\ref{eq:canon}).

For the separation of variables, however, it is enough that the factors
$\D_n(x)$ can be made to depend only on $x_n$ which results in
the separated equations
\beq
 {[}\Xi_n^3-\tau_1(x_n-2\eta)\Xi_n^2+\tau_2(x_n-2\eta)\Xi_n
-d(x_n-2\eta){]}\psi(x_n)=0
\label{eq:sepeq}
\eeq
where $\Xi_n$ stands for the operator
$$ \Xi_n\psi(x_n)=\D_n(x_n)\psi(x_n-\eta) $$

In the $SL(2)$ case the Theorem 3.4 of \cite{Skl:32} establishes the
property in question. The corresponding problem for the $SL(3)$
is being under study.

There is another problem which should be solved before one could
establish the separation of variables in the $SL(3)$ case.
If the common spectrum of the coordinates $\{x_n\}_{n=1}^{3M}$
is a bounded (finite) set
in $\C^{3M}$ and  is not a cartesian product of one-dimensional
sets ${\rm spec}x_n$
then it must have a special geometry for
the separation of variables to take place. The simplest example
is given by the Laplacian in a rectangular triangle with the
zero boundary conditions. The eigenfunctions are not factorized
(\ref{eq:Psi}) but, instead, are linear combinations of such products.

To sum up, the quantum characteristic equation itself provides only local
separation of variables. In order to obtain a global s.\ o.\ v.\
one needs to study more deeply the spectrum of $\{x_n\}_{n=1}^{3M}$
and the representation of the algebra (\ref{eq:xx}), (\ref{eq:XX}),
(\ref{eq:Xx}).
Their properties depend essentially on the representation of the algebra
\T\  taken and we leave the problem for a subsequent study.
It seems to be a plausible conjecture that s.\ o.\ v.\ takes place for
any representation of \T.

\section{Comparison with ABA}

In the $SL(2)$ case the separated equations are known
to be equivalent to Bethe equations defining the spectrum of Hamiltonians
provided the corresponding representation of \T\ has a heighest vector
and thus allows application of the algebraic Bethe ansatz technique
\cite{Skl:32}. It is natural therefore to investigate an analogous
correspondence for the $SL(3)$ case.

The Bethe ansatz for $sl(N)$ case was developed in the papers
\cite{Suth:PRB,KR:JPA,Yang,Kulish:DAN}. We shall use the results of
Kulish and Reshetikhin \cite{KR:LOMI,KR:JPA} who considered the most general
representations of $SL(N)$. As shown in \cite{KR:JPA} the eigenvalues
$\tau_{1,2}(u)$ of $t_{1,2}(u)$ together with the quantum determinant
$d(u)$ can be written down in the form

\begin{eqnarray}
\tau_1(u)&=&\L_1(u)+\L_2(u)+\L_3(u)  \nonumber \\
\tau_2(u)&=&\L_1(u)\L_2(u+\eta)+\L_1(u)\L_3(u+\eta)+\L_2(u)\L_3(u+\eta)
\label{eq:eigv} \\
d(u)&=&\L_1(u)\L_2(u+\eta)\L_3(u+2\eta)
       =d_1(u)d_2(u+\eta)d_3(u+2\eta)  \nonumber
\end{eqnarray}
where the number polynomials $d_{1,2,3}(u)$ are determined by the
parameters of the representation of \T\ in question.

The three ``quantum eigenvalues'' $\L_{1,2,3}(u)$ of $T(u)$
can be expressed in terms of two polynomials $Q_{1,2}(u)$
$$\L_1(u)=d_1(u)\frac{Q_1(u+\eta)}{Q_1(u)}\qquad
  \L_2(u)=d_2(u)\frac{Q_1(u-\eta)}{Q_1(u)}\frac{Q_2(u+\eta)}{Q_2(u)}$$
$$  \L_3(u)=d_3(u)\frac{Q_2(u-\eta)}{Q_2(u)} $$

Eliminating $Q_2(u)$ or $Q_1(u)$ from (\ref{eq:eigv})
one obtains for the polynomials $Q_{1,2}(u)$
the third order finite-difference equations \cite{KR:LOMI}.
\begin{eqnarray}
&& d_2(x-2\eta)d_3(x-\eta)Q_1(x-3\eta)
  -\tau_2(x-2\eta)Q_1(x-2\eta) \nonumber \\
&&\quad +\tau_1(x-\eta)d_1(x-2\eta)Q_1(x-\eta)
   -d_1(x-\eta)d_1(x-2\eta)Q_1(x)
\label{eq:Q1}
\end{eqnarray}
\begin{eqnarray}
&& d_3(x-2\eta)d_3(x-\eta)Q_2(x-3\eta)
  -\tau_1(x-2\eta)d_3(x-\eta)Q_2(x-2\eta) \nonumber \\
&&\quad +\tau_2(x-2\eta)Q_2(x-\eta)
   -d_1(x-2\eta)d_2(x-\eta)Q_2(x)
\label{eq:Q2}
\end{eqnarray}

Making the substitutions
\begin{eqnarray}
   Q_1(x)&=&d_2(x)d_3(x+\eta)d_2(x-\eta)d_3(x)\phi(x) \nonumber \\
   Q_2(x)&=&d_3(x)\psi(x) \nonumber
\end{eqnarray}
and using the shift/mul\-tip\-li\-ca\-tion operators
\begin{eqnarray}
\Theta_1\phi(x)&=&d_2(x-2\eta)d_3(x-\eta)\phi(x-\eta) \nonumber \\
\Theta_2\psi(x)&=&d_3(x-\eta)\psi(x-\eta) \nonumber
\end{eqnarray}

one can put (\ref{eq:Q1}), (\ref{eq:Q2}) into the form
\beq
 {[}\Theta_1^3-\tau_2(x-2\eta)\Theta_1^2+\tau_1(x-\eta)d(x-2\eta)\Theta_1
-d(x-\eta)d(x-2\eta){]}\phi(x)=0
\label{eq:Xi1}
\eeq
\beq
 {[}\Theta_2^3-\tau_1(x-2\eta)\Theta_2^2+\tau_2(x-2\eta)\Theta_2-
     d(x-2\eta){]}\psi(x)=0
\label{eq:Xi2}
\eeq

One notices immediately that the equation (\ref{eq:Xi2})
coincides with the hypothetical separated equation (\ref{eq:sepeq})
giving thus a support to the conjecture.

In \cite{Skl:32}
it was conjectured that the separated coordinates should be splitted
into two subsets producing the two separated equations
(\ref{eq:Xi1}), (\ref{eq:Xi2}) which seems now to be
an overcomplication.
The equation for (\ref{eq:Xi1}) should be considered rather as the
separation equation
for the alternative set of coordinates obtained from the matrix $U(u)$
in the same way as $x_n$ are obtained from $T(u)$.

\bigskip
\noindent{\it Acknowledgments.}
I am grateful to V.\ B.\  Kuznetsov and V.\ O.\ Tarasov
for valuable and encourageing discussions.
I thank the Isaac Newton Institute for Mathematical Sciences
for hospitality.

\end{document}